\documentclass[twocolumn,aps,prl,superscriptaddress,showpacs,secnumroman,showkeys]{revtex4}
\usepackage{amsmath,bm}%
\usepackage{graphicx}%
\usepackage{epsf,epsfig,latexsym}
\usepackage{enumerate}
\usepackage{floatrow}
\begin{document}
\title{An experimental survey of the production of alpha decaying heavy elements in the reactions of $^{238}$U +$^{232}$Th at 7.5-6.1 MeV/nucleon 
}  
\author{S. Wuenschel}
\affiliation{Cyclotron Institute, Texas A$\&$M University, College Station, 
Texas 77843}
\author{K. Hagel}
\affiliation{Cyclotron Institute, Texas A$\&$M University, College Station, 
Texas 77843}
\author{M. Barbui}
\affiliation{Cyclotron Institute, Texas A$\&$M University, College Station, 
Texas 77843}
\author{J. Gauthier}
\affiliation{Cyclotron Institute, Texas A$\&$M University, College Station, 
Texas 77843}
\author{X. G. Cao}
\affiliation{Institute of Modern Physics HIRFL, Chinese Academy of Sciences, 
Lanzhou, 730000, China}
\affiliation{Cyclotron Institute, Texas A$\&$M University, College Station, 
Texas 77843}
\author{R. Wada}
\affiliation{Cyclotron Institute, Texas A$\&$M University, College Station, 
Texas 77843}
\author{E. J. Kim}
\affiliation{Cyclotron Institute, Texas A$\&$M University, College Station, 
Texas 77843}
\affiliation{Division of Science Education, Chonbuk National University, 567 
Baekje-daero Deokjin-gu,Jeonju 54896, Korea
}
\author{Z. Majka}
\affiliation{Smoluchowski Institute of Physics, Jagiellonian University, 
Krakow, Poland}
\author{R. P\l{}aneta}
\affiliation{M. Smoluchowski Institute of Physics, Jagiellonian University, 
Krakow, Poland}
\author{Z. Sosin}
\affiliation{Smoluchowski Institute of Physics, Jagiellonian University, 
Krakow, Poland}
\author{A. Wieloch}
\affiliation{Smoluchowski Institute of Physics, Jagiellonian University, 
Krakow, Poland}
\author{K. Zelga}
\affiliation{Smoluchowski Institute of Physics, Jagiellonian University, 
Krakow, Poland}
\author{S. Kowalski}
\affiliation{Institute of Physics, University of Silesia, 40-007 Katowice, Poland.}
\author{K. Schmidt}
\affiliation{Institute of Physics, University of Silesia, 40-007 Katowice, Poland.}
\author{C. Ma}
\affiliation{Institute of Particle and Nuclear Physics, Henan Normal 
University, Xinxiang 453007, China}
\author{G. Zhang}
\affiliation{Shanghai Institute of Nuclear Research,
Chinese Academy of Sciences, Shanghai 201800, China}
\author{J. B. Natowitz}
\affiliation{Cyclotron Institute, Texas A$\&$M University, College Station, 
Texas 77843}

\date{\today}

\begin{abstract}
The production of alpha particle decaying heavy nuclei in reactions of 
7.5-6.1 MeV/nucleon $^{238}$U +$^{232}$Th has been explored using an in-beam  
detection array composed of YAP scintillators and gas ionization chamber-Si 
telescopes. Comparisons of alpha energies and half-lives for the observed 
products with those of the previously known isotopes and with theoretically 
predicted values indicate the observation of a number of previously 
unreported alpha emitters.  Alpha particle decay energies reaching as 
high as 12 MeV are observed. Many of these are expected to be from 
decay of previously unseen relatively neutron rich products.  While the 
contributions of isomeric states require further exploration and specific 
isotope identifications need to be made, the production of heavy isotopes 
with quite high atomic numbers is suggested by the data.  
\end{abstract}



\maketitle
 
\section*{I. INTRODUCTION}
The synthesis of and the characterization of the properties of heavy and 
super-heavy elements is one of the important current focal points in both 
experimental and theoretical nuclear science. Very high atomic number 
nuclei have long been predicted to exhibit new stabilizing shell structures 
as well as possible exotic shapes such as toroids and bubbles. See 
references~\cite{hermann79,flerov83,armbruster99,greiner99,oganessian06,oganessian15,decharge99,bender01,wong73} 
and those within. Studies of the chemical properties of new heavy elements 
are being employed to establish their chemical families and serve to provide 
stringent new tests of our understanding of relativistic effects in 
electron structure~\cite{pershina07,dullmann02,jungklas13,stakemannxx,hahn82}. 

Model predictions for a shell stabilized ''island of stability'' differ in 
the locus of the center of that island, but agree in their prediction that 
the fission barriers in the island region reduce the probability of 
fission during de-excitation of the primary excited nuclei produced in 
synthesis reactions and mitigate against the spontaneous fission decay 
mode of those 
isotopes~\cite{bender13,kiren12,staszczak13,baran15,agbemava15,agbemava17,anghel17,giuliani17,karpov12,martinez12,marketin16}. Thus the main modes of decay 
in and near these islands are predicted to be alpha and beta 
decay~\cite{kiren12,staszczak13,baran15,karpov12,martinez12,marketin16}.

The synthesis technique which is typically used to search for new heavy 
isotopes is fusion of a heavy target nucleus with a light to medium projectile 
nucleus~\cite{oganessian06,oganessian15,oganessian99,oganessian00,oganessian06_1,hamilton15,hofmann15,utyonkov16,oganessian17}.
The compound nuclei formed  have excitation energies which favor fission 
into two medium mass nuclei rather than gentler sequential emission modes. 
As a result the net production probability for heavy nuclei which survive 
fission usually decreases rapidly with increasing atomic number of the fused
system~\cite{oganessian99,oganessian00,oganessian06_1,hamilton15,hofmann15,utyonkov16,oganessian17}.

Fusion of doubly-magic neutron-rich $^{48}$Ca projectiles with trans-uranium 
target nuclei has led to the synthesis of elements as high as 
Z = 118~\cite{oganessian99,oganessian00,oganessian06_1,hamilton15,hofmann15,utyonkov16,oganessian17}. 
For the reaction used to produce element 118, Oganesson, the reaction 
cross section using $^{48}$Ca is $\sim 0.5$ 
picobarns~\cite{oganessian99,oganessian00,oganessian06_1}. Such 
cross-sections severely limit the prospects for heavy element research. 
Even when the projectiles are neutron rich the compound nuclei produced 
are neutron deficient relative to the line of beta stability.

The limitations of fusion reactions have led to a renewed interest in 
exploring alternative reaction mechanisms for production of neutron rich 
heavy and super-heavy isotopes. In particular considerable theoretical 
effort has been devoted to exploring the use of multi-nucleon transfer 
reactions between pairs of heavy 
nuclei~\cite{zagrebaev13,zagrebaev13_1,tian08,zhao16,karpov17,golabek10,welsh17,kratz15}.  
This technique received some earlier attention from both experimentalists and 
theorists~\cite{schadel16,hildenbrand77,gaggeler80,jungclas78,freisleben79,reidel79,kratzxx,schadelxx,trautmann78,kratz13} 
but, based on the early experimental results was not pursued for heavy 
element synthesis. 

Recent new approaches employed to model the initial multi-nucleon transfer 
stage of such reaction processes typically calculate yields and excitation 
energies of “primary” isotopes and then employ statistical decay models to 
predict the final product distributions resulting from the ensuing 
de-excitation 
stages~\cite{zagrebaev13,zagrebaev13_1,tian08,zhao16,karpov17,golabek10}. 
Fission is, of course, the key competing de-excitation mode which limits 
the heavy isotope survivability and spontaneous fission can compete directly 
with alpha or beta decay. Predicted fission barriers and alpha decay 
energies rely upon model-dependent mass surface 
extrapolations~\cite{kiren12,staszczak13,baran15,agbemava15,agbemava17,anghel17,giuliani17,karpov12,martinez12,marketin16}. 
The predicted survival cross sections for heavy and super-heavy nuclei 
are extremely sensitive to details of these mass surface extrapolations 
and the location of closed shells. Uncertainties of 1 MeV in the fission 
barriers can lead to an order of magnitude change in the fission 
probabilities. Uncertainties in level densities, temperature dependencies 
of fission barriers and details of the fission dynamics further complicate 
calculations of fission probabilities. While quantitative predictions vary 
widely, systematic theoretical studies of survival probabilities carried 
out using both statistical models and microscopic model calculations of 
fission rates indicate high survival probabilities in and near the 
island of 
stability~\cite{kiren12,staszczak13,baran15,anghel17,giuliani17,karpov12,martinez12,marketin16}. 
Notably, recent microscopic fission model results indicate significant 
increases in fission survivability compared  to those of statistical models 
employing the same fission barriers~\cite{zhu17,xia11}.  Indeed, a strong 
increase in survivability is already evident in the experimental fusion 
cross section data for the heaviest 
elements~\cite{hamilton15,hofmann15,utyonkov16,oganessian17}. 

Some calculations suggest that near the valley of stability, beta decay 
competes with alpha and fission decay and that short-lifetime beta minus decay 
will be dominant for the more neutron rich isotopes in that 
region~\cite{karpov12,martinez12,marketin16}.  This raises the interesting 
possibility that the production of neutron rich lower Z products can feed 
higher Z products through $\beta^-$ decay, increasing the effective production 
cross section for such higher Z products near the line of stability. 
Recent systematic efforts to explore the utility of multi-nucleon transfer 
reactions for production of new neutron-rich isotopes suggest that the 
experimental cross sections exceed predicted cross 
sections~\cite{welsh17,kratz15}. It is interesting to ask whether a 
similar trend exists for heavier elements.  Good experimental data are
 needed to guide future efforts in heavy element research. 

\section{Experimental Procedure}
In some earlier work on this problem we used the BigSol 
Superconducting-Solenoid Time of Flight Spectrometer at Texas A \& M to 
perform several surveys of projectile target combination and bombarding 
energy for collisions of $^{86}$Kr, $^{136}$Xe and $^{197}$Au with $^{232}$Th 
in an effort to identify good candidate reactions for heavy and super-heavy 
element production~\cite{odonnell99,barbui06,barbui10,majka14,andrej}. 
Those experiments, at higher laboratory energies per nucleon than the 
present work, indicated the possible production of heavy elements with Z 
above 100~\cite{andrej}. However the experiment was discontinued when the 
spectrometer developed a He leak which made it not possible to sustain the 
necessary magnetic field.  We then adopted a new direction for investigation 
of such reactions based upon the implantation of heavy reaction products 
in a down-stream catcher foil and the detection of alpha particle decays 
characteristic of heavy nuclei.  For this purpose the Jagellonian University 
Group constructed a forward array of 63 active catcher (AC) fast plastic 
scintillator detectors and dedicated state-of-the-art fast timing electronics 
to function as a time filter for recoil implantation and alpha decay 
detection~\cite{majka14,andrej}.  Tests employing these plastic 
scintillators demonstrated that the use of such a time filtering device was 
feasible even in the harsh environment encountered in the experiments 
envisaged.  The test experiments indicated a possible production of alpha 
decaying heavy elements.  However, while the fast plastics provided optimum 
time resolution, the quenching of the light-output inherent in solid 
scintillators and the inability to do pulse shape discrimination with 
the plastic meant that discrimination between high energy alpha particles and 
spontaneous-fission fragments was difficult.

\begin{figure}[b]
\epsfig{file=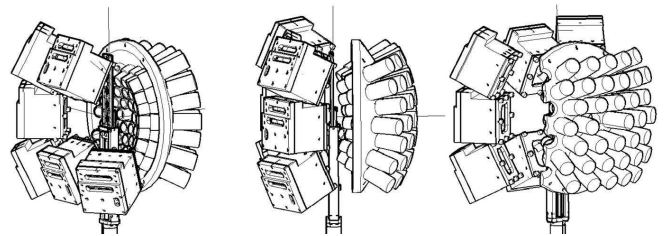,width=9.2cm,angle=0}
 \caption{Schematic diagram of IC-Si Detectors and YAP active catcher array. 
The three views are from three different angles. In the central view the beam 
enters from the left.}
\label{fig1}
\end{figure}

\begin{figure}
\epsfig{file=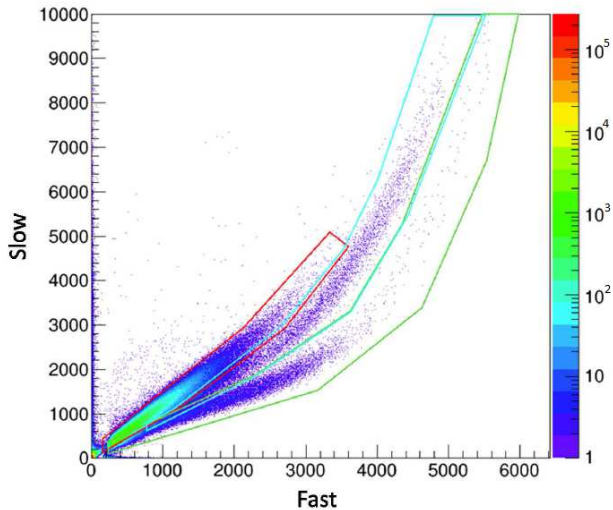,width=8.5cm,angle=0}
 \caption{Pulse-shape discrimination: Amplitude of slow portion of the AC 
signal vs amplitude of the fast portion (peak) signal.  Windows indicated 
are, from left to right, alpha partiucles, fission fragments, beam and heavy 
recoils.  Data for both beam-on and beam-off are included. }
\label{fig2}
\end{figure}

Therefore, to carry out the present experiments we constructed an active 
catcher system consisting of a 40 detector array of yttrium aluminum 
perovskite, YAP,  scintillators coupled to Hamamatsu photo-multiplier 
tubes, PMT, via Lucite light guides. See Figure~\ref{fig1}.  The YAP 
scintillators were chosen because of the fast rise time and light decay 
properties ($t_1\sim 14$ns, $t_2\sim 140$ns) that provide access to particle 
identification through pulse shape discrimination.  This capability is 
employed to distinguish between alpha decay and fission fragments or degraded 
beam and recoiling reaction products.  This is important because the 
non-linear response of the solid YAP scintillator makes energy signals alone 
insufficient for complete separation.  The particle identification is 
demonstrated in figure~\ref{fig2} where we plot the slow component of the 
light versus the fast component.  The gates for the different identified 
products are shown on the plot.  The PMTs were powered by custom made 
active bases.  The active bases provide the capacity to 
handle $\sim 100\times$ more events/second than the Hamamatsu pasive bases
before PMT gain sagging becomes an issue~\cite{ren17}.  This resulted in 
additional beam intensity capacity.  During offline testing, the active 
catcher modules (YAP-light guide-PMT) exhibited $<10$\% resolution for the 
8.78 MeV alpha-decay peak of $^{228}$Th. 
In the experiment the array had a total geometric efficiency of 22\% for 
forward-recoiling products in the angular range of 7 to 60 degrees.  As 
noted below the experiments reported in this paper were carried out in a 
pulsed beam mode.  During beam-off the intrinsic detection efficiency for 
alpha decays in the AC was $>50$\% (depending on implantation 
depth).  As employed the AC array was sensitive to products 
with transit times of only a few nanoseconds (much shorter than those of 
spectrometer experiments) originating from various reaction mechanisms. 
This array was employed with a backward array of gas ionization 
chamber-silicon telescopes (IC-Si) capable of detecting alpha particles 
emerging from the forward catcher. See Figure~\ref{fig1}.
 
An annular ring shielded the IC-Si telescopes from emission from the target. 
This IC-Si telescope array, active in both beam-on and 
beam-off modes had an overall geometric efficiency of 6\% for alphas 
originating in the active catcher and an $\alpha$-particle identification 
threshold of 5.6 MeV. In addition to providing detection and identification 
of the alpha particles emerging from the YAP array, the coincidence capability 
thus realized provides a reconstruction of total alpha energy for those 
emerging alpha particles detected in the backward direction as well as 
information on implantation depth. The SRIM range-energy code was  
used to derive the required range-energy information for the implantation 
depth calculations~\cite{zeigler10}. In the experiment implantation 
depths of 2 to 22 microns were observed for accepted coincident alpha 
particles. These particles had total energies as large as 12 MeV.  Some 
apparently higher energy $\alpha$-particles were observed by the IC-Si 
detector.  These had unphysical apparent depths and were attributed to long 
range alphas from ternary fission with attendant larger AC coincidence 
energies resulting from simultaneous detection of fission fragments.

During the experiment one of the IC-Si detectors was blinded by a thick 
degrader. This allowed us to evaluate possible spurious events which 
might arise from ($n,\alpha$) reactions in the detector materials.  This 
effect was found to be negligible, consistent with GEANT simulations of this 
possibility~\cite{geant}. 

The time decay constants inherent in YAP scintillators are notably slower 
than the fast plastic utilized initially. Thus, the dedicated, custom-made 
electronics and trigger scheme employed for the plastic scintillator array 
could not be easily adapted to these detectors.  For this reason we turned 
to commercially available electronics for the YAP array. An experimental 
set-up employing a triggering and signal acquisition scheme based upon the 
Struck SIS3316 250MHz Flash ADC modules was developed. These modules 
provide flexible digital triggering mechanisms. 

Although the direct catcher technique does require us to work in a rather 
hostile environment, it has an advantage relative to the spectrometer 
in the much shorter transit times of the recoils (a few nanoseconds) 
which means that activities with much shorter lifetimes can be investigated. 
We emphasize that the present experiment was intended to provide a broad 
based survey and could be followed up by more targeted experiments guided 
by these results.

In July 2016, experimental data were taken using the YAP active catcher 
array coupled to the backward angle IC-Si detector modules. Beams of 
$^{197}$Au and $^{238}$U of 7.5 MeV/nucleon were incident on 11 mg/cm$^2$ 
$^{232}$Th targets.  The beam emerged from this target with an energy well 
below the coulomb barrier of 6.1 MeV/u. 

The trigger scheme employed in these experiments was based on three 
operational considerations. 
\begin{enumerate}
\item The experiment could be carried out in a pulsed beam mode with 
variable beam-on/beam off times.
 
\item The backward angle silicon detector modules generate triggers at a 
relatively low rate and very high quality. 

\item  Vetoing beam-on signals with the RF signal would have allowed the 
SIS3316 modules to trigger in a mode very similar to the Jagellonian 
University analog electronics.  However, since the RF signal is about 
5ns wide, the Flash ADC bins are 4ns wide and the YAP signals are about 
5ns wide, the convolution of these signals did not allow to trigger closer 
than 17ns from the RF signal which meant that such operation would have 
required vetoing about 30\% of the time.
\end{enumerate}
 
To avoid the problems associated with point 3, we decided, in this 
experiment, to allow the forward angle YAP detectors to trigger acquisition 
only during the beam-off periods.  

Triggering of the acquisition utilized two primary modes, beam-on and 
beam-off.  During the beam-on periods, only the silicon detectors triggered 
the acquisition.  The active catcher array was read in slave mode and 
waveforms were stored for $2 \mu s$ for each active catcher module. The 
synchronization between Si and YAP was set so that a coincident peak in 
an active catcher module would appear at $\sim 790$ ns into the $2 \mu$s 
flash ADC storage period.  During the beam-off periods, the active catcher 
detectors were permitted to trigger the acquisition.  Waveforms were stored 
only for modules that triggered during the event. Because the trigger was 
generated entirely digitally, the beam-on/off trigger mode was swapped 
using beam-on/off bits provided to the acquisition system. During this 
experiment two different pulsing patterns were employed; 100 ms on/ 
30 ms off and 30 ms on/30 ms off. 

A third overarching trigger was also built into the logic. This 
intermittent trigger was applied to the silicon detectors. The SIS3316
 modules have a binary threshold mode.  The secondary threshold can be 
used to either veto an event, or as in our case, generate a secondary 
logic signal routed to another lemo output.  The threshold for this 
trigger was set to 8-8.5 MeV energy in the silicon detectors. Following 
an event generating this second, “high energy trigger” signal, the beam 
was completely turned off for 20 seconds and the acquisition set into the
 beam-off trigger mode.  Additionally, for such events, the flash ADC storage 
periods were extended to be $160 \mu$s long. 

Using the multiple trigger modes it was possible to efficiently  explore 
alpha spectra during beam-off periods of $2 \mu$s, $160 \mu$s, 
30 ms and 20 seconds and beam-on periods of 100 ms and 
30 ms.  

Our original intention for beam monitoring for cross section 
determinations was to use active catchers at larger angles to directly 
count elastically scattered particles.  The change in triggering for the 
YAP detectors prevented this so beam monitoring was done using a Faraday 
cup in the fringe field at the exit port of the accelerator.

For the data analysis an offline peak finding algorithm was developed 
based on the trapezoidal digital filter used in the SIS3316 triggering
 process.  The response of this algorithm also generated the fast portion 
of the pulse shape discrimination.  A minimum of 20-40 ns separation results 
from the settings chosen for this algorithm which were optimized for YAP 
and the 4 ns buckets of the FADC.  Currently, deconvolution of peak pile-up 
is not built into the analysis package.  This creates an effective minimum 
distance between particle identified peaks of approximately 80-100 ns.  
Tests with 40 ns minimum distances revealed little change in acceptance 
rates.  Pile-up of pulses separated by less than 16 ns could result in 
errors in derived peak energies.  Visual inspection of high energy peaks 
of interest was employed to exclude this possibility. 

\begin{figure}[b]
\epsfig{file=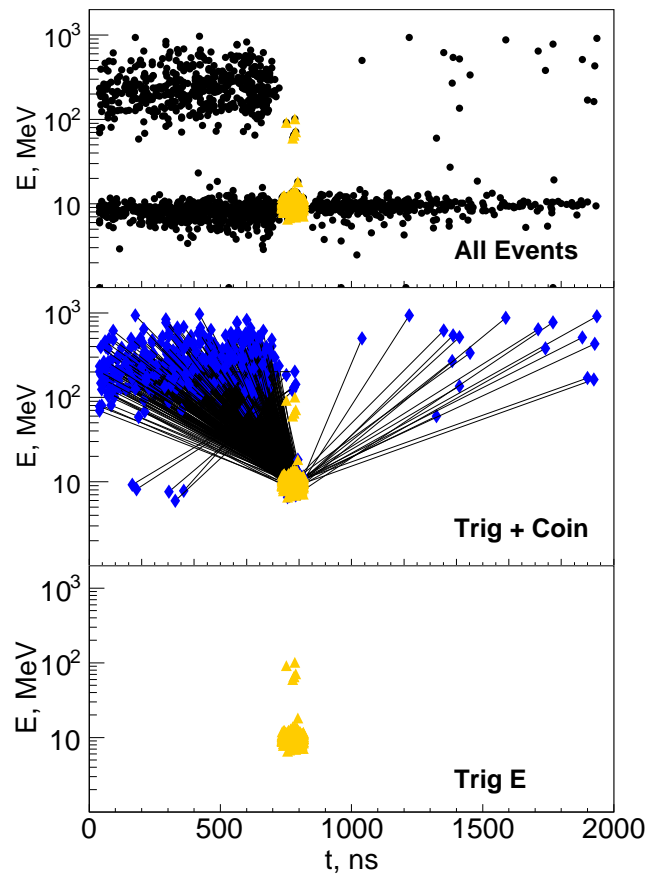,width=8.5cm,angle=0}  
 \caption{Recorded energies and times for IC-Si triggered events having 
more than one AC signal in the $2 \mu$s Flash ADC inspection time.  Events 
are for detectors in the angular range of 30 to 50 degrees. IC-Si triggers 
appear at $\sim 790$ ns. Lines indicate the correspondence between alpha 
particle signals and heavy product signals observed in the same $2 \mu$s 
period. Top- all events; Middle- coincident heavy product-alpha particle 
events; Bottom- trigger events.  See text.}
\label{fig3}
\end{figure}

\section{Analysis and Experimental Results}
As a first result from the experiment, we present in Figure~\ref{fig3}, 
a plot of assigned energies (AC or IC+ Si+AC) for events in which more 
than one flash ADC signal was registered in the $2 \mu$s recording window 
associated with an IC-Si trigger. 
 
The actual trigger signals appear at $\sim 790$ ns in this plot.  While most 
of these events have one other peak, some have two. Thus in the plot we see 
energies of other alpha particles detected in the AC during the inspection 
period. We also see a number of much higher energy signals.  The overwhelming 
number of these signals precede the trigger. In the figure we have also 
included lines connecting each of these high energy signals to alpha 
particle signals seen in the same AC module during the same $2 \mu$s 
Flash ADC recording period. We conclude that these signals correspond to 
the implantation of heavy alpha decaying recoils which precede the trigger 
decay within the $2 \mu$s window. Further confirmation of this is that we 
have observed target to catcher flight times and recoil energy-implantation 
depth correlations (derived from the energy loss of the alpha particles 
emerging from the AC) which are consistent with the energies assigned. 
The recorded recoil to trigger flight times are employed in a later 
section to determine apparent half-lives for this subset of events.

We also note in figure~\ref{fig3} several trigger events with total 
energies $\sim 100$ MeV.  These events correspond to detection of an 
identified alpha particle in the IC-Si associated with a signal 
identified as fission in the AC (The fission energy calibration is 
only approximate).  These appear to correspond to ternary fission events 
emitting long range alpha particles~\cite{geant}. 

\begin{figure}
\epsfig{file=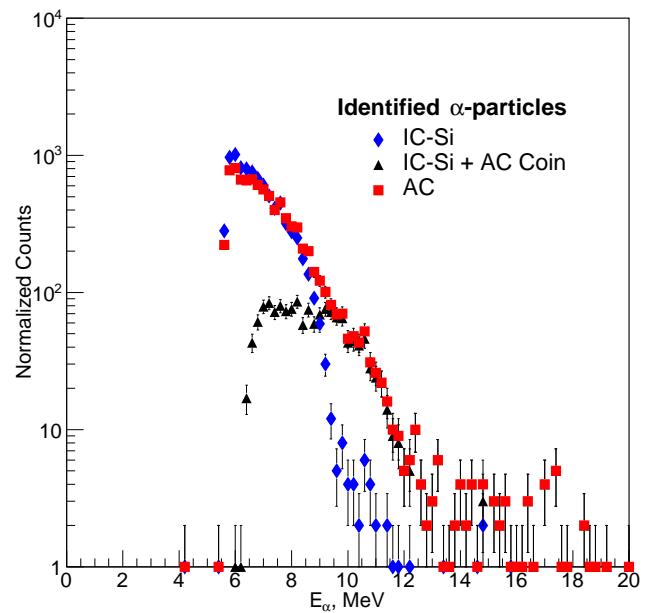,width=8.5cm,angle=0}
 \caption{Comparison of alpha particle kinetic energy spectra. Diamonds- 
EnergySpectrum in IC-Si telescope including window correction; 
Triangles-Energy Spectrum Sum of IC-Si energy plus coincident AC energy; 
Squares - Energy spectrum in AC detector.  The last two spectra are 
normalized at 10 MeV.}
\label{fig4}
\end{figure}

We present in Figure~\ref{fig4}, a comparison between energy spectra of the 
Si-IC detected events (including a window correction), of the AC detected 
events and of the combined IC-Si-AC detected events. The agreement between the 
last two is very good, providing important confirmations of the individual 
detector calibrations and the pulse shape identification techniques 
employed to identify alpha particles in the YAP detectors. 
 
A careful exploration of the IC-Si trigger events using their apparent 
implantation depths indicated that identified alpha particles with total
energies above $\sim 11.5$ MeV corresponded to alpha emission in ternary 
fission events or included possibly misidentified YAP signals at the limit 
of our pulse shape discrimination capabilities. Therefore in the analyses 
reported below we have limited ourselves to identified alpha particles with 
energies $\leq 11.5$ MeV.  

\begin{figure*}
\epsfig{file=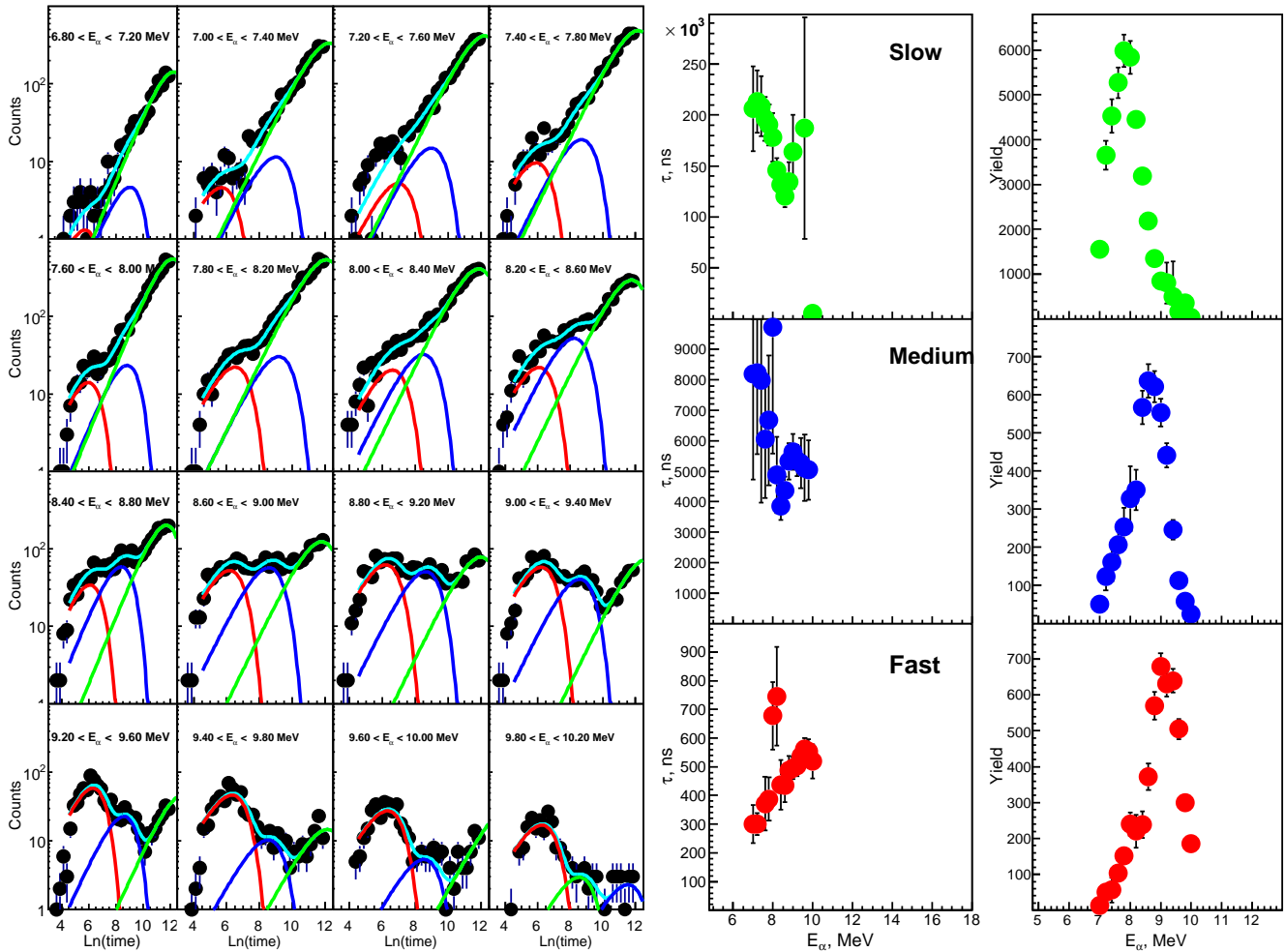,width=18cm,angle=0}
 \caption{Example of fitting process results.  On left the results of 
fitting the function of equation (1) to $\alpha$-energy selected data 
from the millisecond pulsing range are shown. Data-solid circles,
 Source fits-lines.  On the right are derived mean lifetimes and yields from 
the three component fits versus the average of the selected bin. 
(colored on-line)}
\label{fig5}
\end{figure*}

The resolution of the YAP detectors is such that resolving emission from 
individual isotopes in the midst of the large number of isotopes with 
similar alpha decay energies is extremely difficult. Thus we have instead 
elected to explore overlapping sequential bins of alpha energy, 400 keV 
wide, displaced each time by 200 keV to survey the dominant decay times as 
a function of energy. These fits were restricted to average energies 
below 11.5 MeV, based upon the implantation depth information described above.

For each energy bin we employed the method suggested by 
K. H. Schmidt \textit{et al.} to explore decay time
distributions~\cite{mutterer08}.  For a given decaying nucleus the decay 
time distribution data is characterized by a universal function.  In the 
fitting parent daughter relationships which exist are not explicitly taken 
into account.  This, and the limitations of the three source assumption 
mean that the fit results are primarily indicative of the decay times 
of the nuclei whose yields are dominant in a sampled energy range.  We 
return to the question of parent-daughter relationships later in this paper.
The universal function is given by
\begin{equation}
\frac{dn}{d\theta} = n\lambda e^{\theta}e^{-\lambda e^{\theta}}
\end{equation}
in which $\theta = \ln{t}$ where $t$ is the decay time and the free 
parameters are $n$, the total number of counts and $\lambda=1/\tau$ 
where $\tau$ is the mean life time. The most probable value of this 
distribution is $\ln{\tau}$. We have employed this function as a fitting 
function to explore the decay curves as a function of alpha energy in each 
time region. Implicit in this approach is that the times are generally 
measured from the beginning of the decay period explored. However in the 
particular case of the $2 \mu$s and $160 \mu$s flash ADC recording periods, 
we have observed recoil-alpha-patrticle coincidences.  For such events the 
times are those between recoil and alpha-particle detection.  Corrections 
for recoil and alpha flight times differences are small. 

\begin{figure*}
\epsfig{file=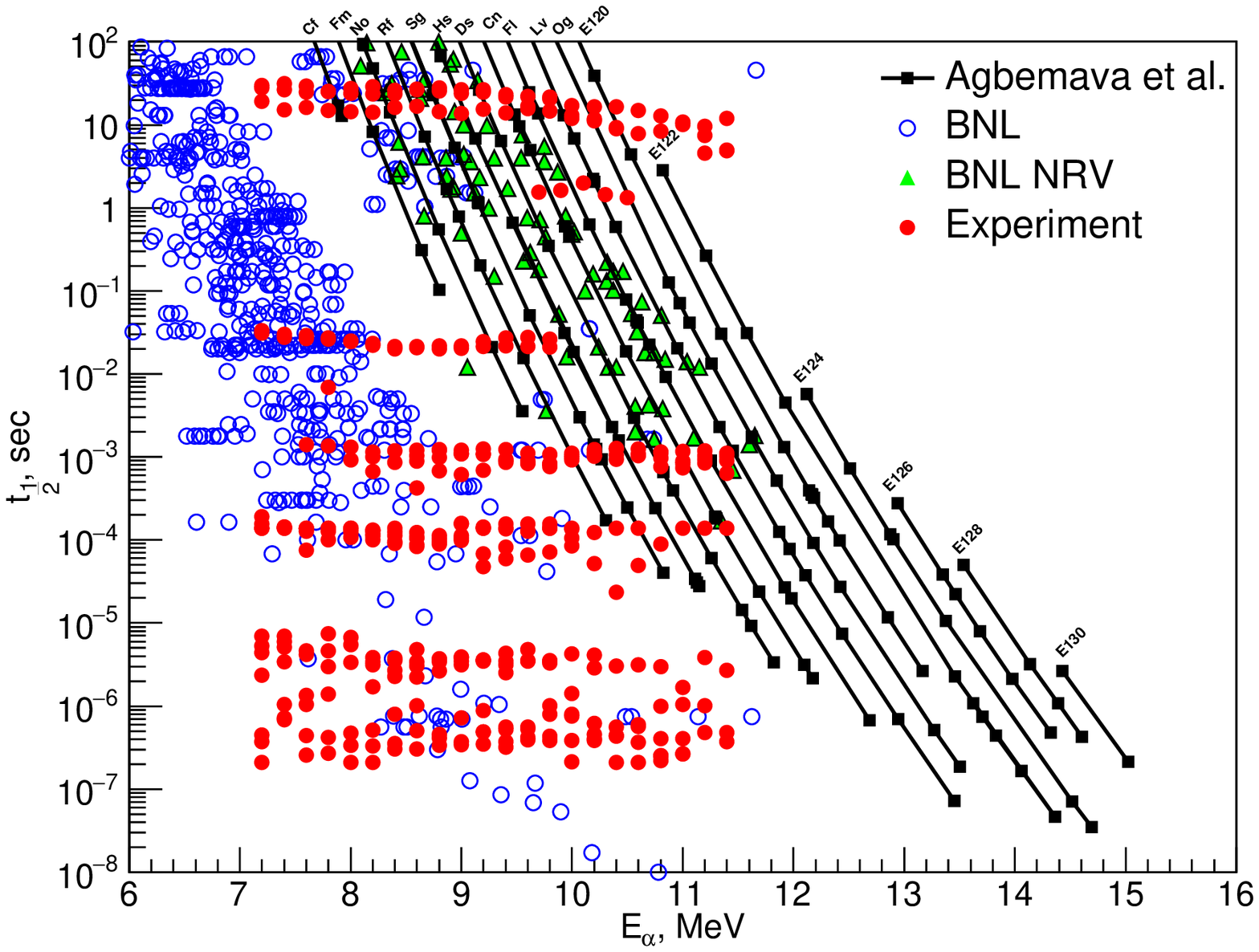,width=19cm,angle=0}
 \caption{Grand summary. Total Half-life, seconds  vs E$_\alpha$, MeV for 
activities with t$_{1/2}$ $\leq$ 100 seconds, Open circles- known alpha 
half-lives for isotopes with Z to 101(BNL table), solid diamonds- NRV and BNL 
data for isotopes  with Z $>101$, solid black squares connected by lines 
depict predictions of partial alpha-lifetimes vs E$_\alpha$ for even-even 
nuclei, left to right Z=98 to 130. Data for different time ranges are 
represented by closed circles (see text) (color online).}
\label{fig6}
\end{figure*}

Figure~\ref{fig5} shows an example of the fitting strategy pursued. In that 
figure the results of 3-source fitting for bins of mean energy ranging 
from 6.8 to 10.2 MeV are shown.  The three sources are qualitatively 
identified as fast, medium and slow.  The derived values of the mean 
lifetimes, $\tau$, and normalization constants, n, are plotted as a function 
of mean energy. 

In Figure~\ref{fig6}, we summarize the results of this investigation, plotting 
half-life in seconds vs the alpha particle kinetic energy in MeV. The 
apparent clustering into seven dominant time ranges reflects weighted 
averages of the activities falling within the selected alpha particle 
energy windows for the three-source approximation and the pulsing protocol
chosen.  For later discussion, we identify these groups as group 1 - group 7
in order of decreasing half-life range (top to bottom).

\begin{figure}
\epsfig{file=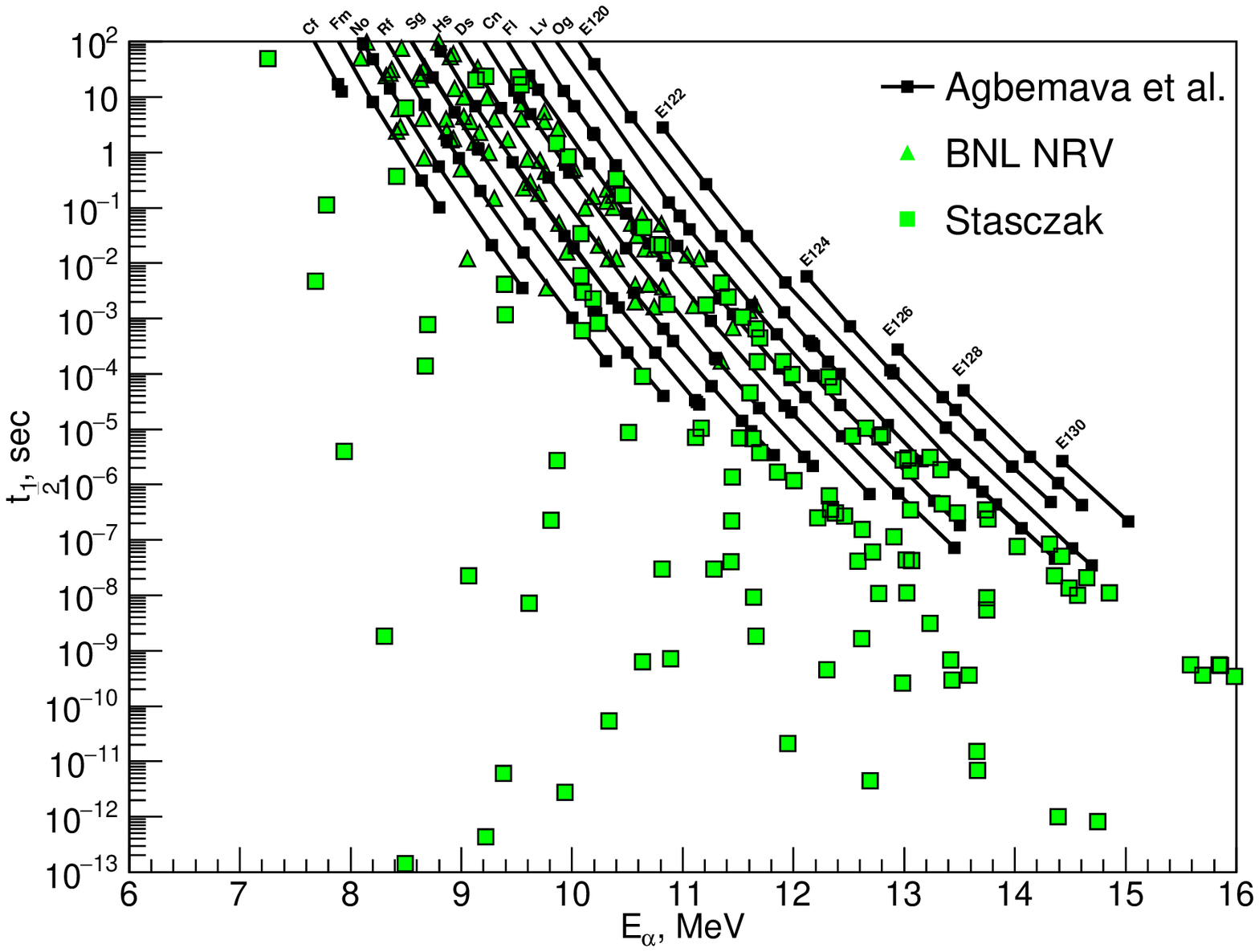,width=9.2cm,angle=0}
 \caption{Comparison of the predicted partial alpha decay half-lives for 
E-E heavy elements~\cite{agbemava15} with predicted total half-lives 
including  spontaneous fission decay~\cite{staszczak13}.}
\label{fig7}
\end{figure}

For comparison to the data we present three other sets of information. 
The first set, indicated by open circles, represents the experimental 
data for t$_{1/2}$ vs alpha energy for previously identified alpha-decaying 
isotopes with Z $\leq 101$~\cite{bnl}.  The second, indicated by closed
triangles represents the existing experimental data for elements heavier 
than 101~\cite{bnl,nrv}.  The third set, represented by solid squares 
connected by  lines, indicates the values calculated for partial alpha 
decay half-lives for even-even isotopes with Z from 98 (Cf) to 
130-(left to right) and N from 172 to 196 using a density-functional 
approach with the PCPK3 interaction~\cite{agbemava15}.  As is commonly done, 
the authors calculated these partial half-lives employing the usual 
Viola-Seaborg approach with parameters determined from fits to the
 known isotopes~\cite{viola66}. 

Various predictions for the branching ratios for the decay of the heaviest 
of the elements in the region of the valley of stability strongly 
favor $\alpha$ emission~\cite{staszczak13,baran15,karpov12}. Significant 
contributions from other decay modes would lead to smaller total half-lives 
for the nuclei considered. For Even-Odd (E-O), Odd-Even (O-E) and 
Odd-Odd (O-O) nuclei traditionally invoked hindrance factors 
for $\alpha$-decay would lead to some increases in the partial alpha decay 
half-lives compared to those of  the neighboring E-E isotopes~\cite{kiren12}. 

Theoretical calculations of fission barriers and fission lifetimes have 
also been carried out for heavy and super-heavy 
elements~\cite{staszczak13,baran15,agbemava17,anghel17,giuliani17,karpov12}. 
In Reference~\cite{staszczak13} Staszczak \textit{et al.} have calculated 
both alpha decay and spontaneous fission lifetimes for a similar, but more 
limited, set of even-even heavy nuclei than considered in 
reference~\cite{agbemava15}.  In Figure~\ref{fig7}, the total half-life 
predictions of Staszczak \textit{et al.}~\cite{staszczak13} are compared to 
the partial alpha decay lifetimes of 
Agbemava \textit{et al.}~\cite{agbemava15}.  The fission competition included 
in the first often leads to large (many orders of magnitude) reductions in 
the predicted lifetime. The largest changes are in the 8-10 MeV energy 
region, reflecting larger predicted branching ratios for spontaneous fission.  
A number of sub nano-second activities are predicted. Given significant 
branching ratios for spontaneous fission it is possible that the 
experimentally observed sub-millisecond activities in Figure~\ref{fig7} 
correspond to higher Z isotopes than the comparison to partial alpha 
half-lives alone would suggest.
 
The data in Figure~\ref{fig6} indicate the observation of a number of 
previously unreported alpha emitters with energies reaching as high as 
11.5 MeV.  Given the multi-nucleon transfer mechanism in play many of 
these are expected to be previously unseen neutron rich products.  The 
raw comparison between data and predictions in the millisecond and second 
time-ranges shows $\alpha$-particle energies which might represent decay 
from very high Z isotopes.  However, we must recognize that alpha-particles 
emitted from new isomeric states can have energies quite different from 
those of their ground state counterparts and thus would lead to a different 
t$_{1/2}$ energy correlation.  This is well established in the Fr-At region, 
for example~\cite{bnl}.  

\begin{figure}
\epsfig{file=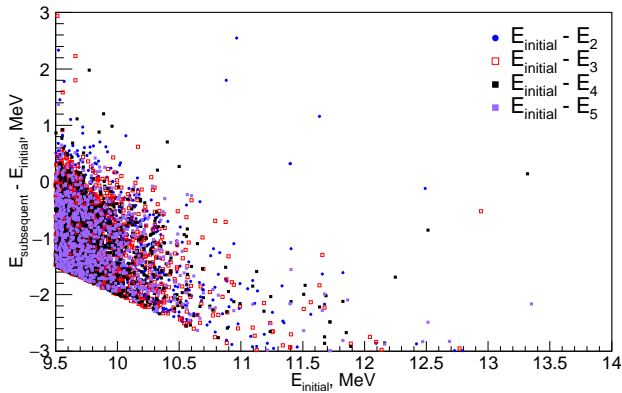,width=8.5cm,angle=0}
 \caption{For 20 second beam-off data- energies of alpha particles emitted 
following an initial alpha particle with E$\geq$9.5 MeV. See text.}
\label{fig8}
\end{figure}

Although the experimental alpha energy resolution (FWHM$ \sim 600$ KeV) 
coupled with the high decay rates observed make searching for individual 
decay chains difficult, we can make an initial test of the isomer 
hypothesis by asking, on an event by event basis, what energies are 
observed following emission of an initial alpha particle of ever increasing 
energy.  For events in which the beam was turned off for 20 seconds we 
present, in figure~\ref{fig8}, the energy differences 
(E$_{subsequent}$-E$_{initial}$) vs E$_{initial}$ where E$_{initial}$ is the 
energy of the first alpha detected in the event and E$_{subsequent}$ are the 
energies for the next 4 alpha particles detected in the same active catcher 
module.  A lower threshold of 9.5 MeV has been imposed on the 
initial $\alpha$-particle energies used for this search.

Up to $\sim 10.6$ MeV initial energy  the observed energy differences span 
an energy range of about 2 MeV and include particles with energies within 
$\sim 0.5$ MeV of the initial energy.  At higher energies the band narrows 
and by an initial energy of 11 MeV most subsequent alpha particles have 
energies more than 1.5 MeV lower.  This is generally larger than predicted 
(and observed) differences in energies of successive ground state decays. 
The populating of alpha decaying isomeric states could explain this 
observation.  Near 11.5 MeV initial energy single events with subsequent 
energies 1.2 and 1.4 MeV lower than the initial energy bear further 
investigation.  Of course isomeric states can also contribute at lower 
decay energies.  To determine the actual identities of the high $\alpha$ 
energy emitters and resolve the question of isomer contributions to our 
spectra requires that detailed decay chain relationships be established. 

\section{Parent-Daughter Relationships}
\begin{figure}
\epsfig{file=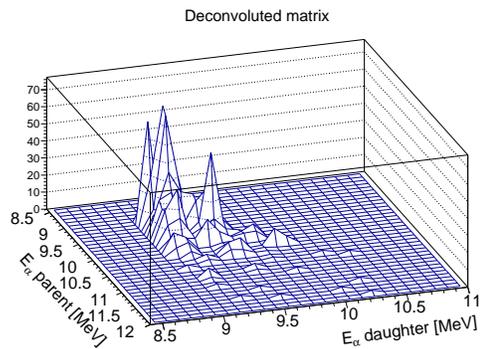,width=7.0cm,angle=0}
 \caption{2D energy-energy correlation plot for data taken during 20 second beam-off periods.}
\label{fig9}
\end{figure}

We have attempted searching for parent-daughter relationships by applying
energy-energy correlation methods analogous to those used in gamma-decay 
spectroscopy~\cite{lee90}.  Two powerful peak searching software packages 
were employed~\cite{root,r}.  As previously noted, and emphasized by the 
correlation plot shown in Figure~\ref{fig9}, the high rates of alpha decay in 
a single AC module coupled with the energy resolution of the present 
experiment make peak searching difficult. Improvements in detector 
resolution and granularity would greatly improve the peak search capabilities.
                   
\begin{table*}[t]
\begin{tabular}{|c|c|c|c|c|}
\hline
 \multicolumn{3}{|c}{Alpha Emission} & \multicolumn{2}{c|}{Spontaneous Fission} \\
\hline
Parent $\alpha$ energy & Daughter $\alpha$ energy & t$_{1/2}$ Daughter & Parent $\alpha$ energy & t$_{1/2}$ Fission \\
  MeV       &    MeV        & sec  & MeV       &sec \\
\hline
9.29  &	9.12 &$1.49\pm 0.32$ &8.15  &$1.86 \pm 0.28$ \\
9.63  & 9.45 &$1.16\pm 0.36$ &8.45  &$1.28 \pm 0.17$ \\
9.75  & 9.12 &$1.35\pm 0.38$ &8.97  &$0.74 \pm 0.35$\\
9.88  & 9.72 &$1.20\pm 0.21$ &9.19  &$1.22 \pm 0.27$ \\
9.92  & 9.36 &$0.96\pm 0.26$ &9.45  &$2.18 \pm 0.37$ \\
10.04 & 9.09 &$0.99\pm 0.55$ &10.05 &$1.83 \pm 1.08$ \\
10.14 & 9.88 &$0.99\pm 0.32$ &      &      \\
10.26 & 9.51 &$1.13\pm 1.18$ &      &     \  \\
\hline
\multicolumn{5}{l}{1.	Search with $\pm 0.15$ MeV standard deviation on $\alpha$ energies} \\
\end{tabular}
\caption{Correlated Pair Half-lives$^1$}
\end{table*}

Nevertheless, during these attempts we did isolate, for the 20 second 
beam-off events, some statistically significant correlated emission pairs 
indicating parent- daughter relationships. Half-lives for the daughters 
could be determined from the measured time differences.

\begin{figure}[b]
\epsfig{file=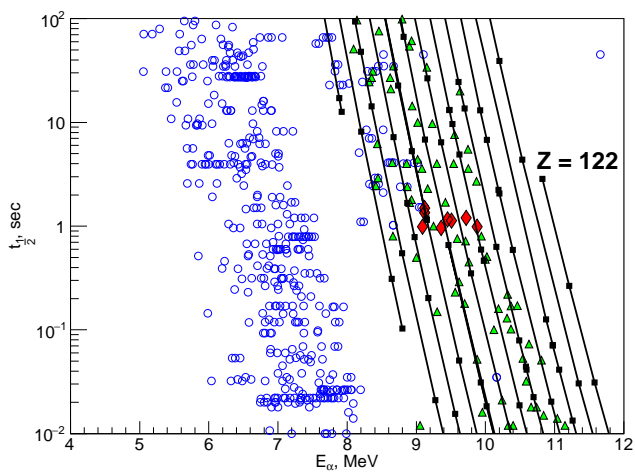,width=9.2cm,angle=0}
 \caption{Experimental results of correlated pair search, solid diamonds. 
For comparison data from previous experiments (Z $\leq$101-solid circles 
and Z $>$101 solid squares) are also indicated, as are the predictions of 
Agbemava \textit{et al.}~\cite{agbemava15} for E-E nuclei with 
Z = 98 to 122( other symbols and lines) }
\label{fig10}
\end{figure}

Half-lives in the 1 to 2 second range are observed for alpha particle 
kinetic energies of 9.3 to 10.3 MeV. These results are presented in Table 1.  
In Figure~\ref{fig10} they are compared with previously reported literature 
results~\cite{schmidt84,bnl}  and with the theoretical predictions for 
even-even nuclei from reference~\cite{agbemava15}.
      
While theoretical predictions for Q$_\alpha$ and t$_{1/2}$ for a specific 
super-heavy isotope vary significantly~\cite{staszczak13,baran15,agbemava15}, 
the phenomenological trends for fixed atomic number, based on the 
Viola-Seaborg-approach~\cite{viola66} and represented by the lines for 
even-even nuclei in Figure~\ref{fig10}, are quite robust.  The comparison to 
the theoretical results suggests that, if these emitters are even-even nuclei, 
they are in a range of Z from 106 to 114.  Recall that these are the 
daughter nuclei in the correlated alpha-particle pairs. The parent nuclei 
would have atomic numbers 2 units higher than the daughters. For even-odd, 
odd-even and odd-odd nuclei the inclusion of phenomenological hindrance 
factors leads to predicted half-lives $\sim 2$ to 10 times longer than those 
for E-E nuclei of the same atomic number. Thus further information is 
required to make definitive atomic number and isotope identifications. 

\section{Spontaneous Fission}
The same energy-energy correlation techniques used to search for alpha-alpha 
correlations were used to search for spontaneous fission decays following 
alpha emission. In this search we identified some alpha-fission correlated 
pairs with parent alpha energies ranging from 8.15 to 10.1 MeV. The 
spontaneous fission daughter half-lives were also found to be in the few 
second range. These results are also summarized in  Table 1. 

\section{Cross Sections}
\begin{figure}
\epsfig{file=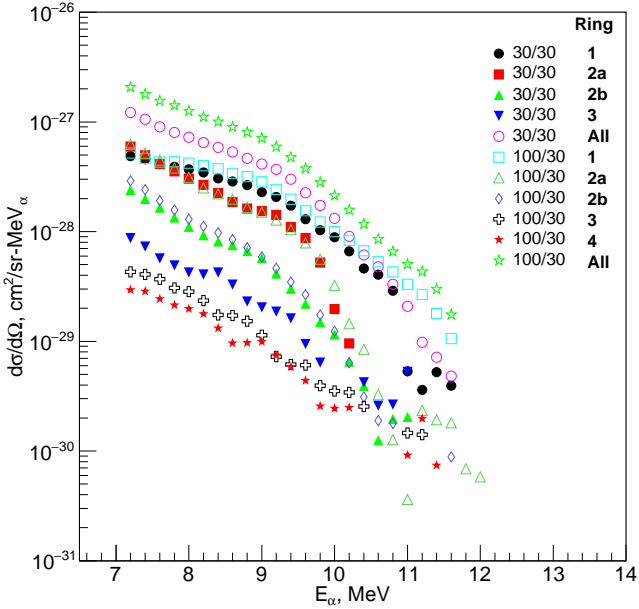,width=8.5cm,angle=0}
 \caption{Integral thick target differential cross sections 
cm$^2$ / sr-MeV vs alpha particle energy.  Angular ranges are 
Ring 1 16$^o$-30$^o$, Ring 2a 31$^o$ -45$^o$, Ring 2b 36$^o$ -50$^o$, Ring 3,
 47$^o$ -51$^o$, Ring 4 52$^o$ -66$^o$.}
\label{fig11}
\end{figure}

To determine cross sections from the three source fit results we have 
assumed that a secular equilibrium with the beam is achieved for each 
activity which is short relative to the relevant pulsing time. In this 
case the normalization constant of the fitting function is the number of 
nuclei present when the beam is turned off (integrated over the number of 
pulsing cycles).  With the secular equilibrium assumption the cross 
sections are easily derived.  In Figure~\ref{fig11} we show, thick target 
differential cross sections as a function of alpha particle energy for 
the 20 sec beam-off events in group 1.

It is important to emphasize that these average cross sections for these 
alpha energy ranges are derived from integral thick target production rates 
assuming that the entire energy range from incident beam energy down to 
the Coulomb barrier is contributing. They include all feeding from parent 
activities during the irradiation. In addition, the energy resolution is 
such that  more than one isotope will contribute in the selected energy 
windows.

The strong decrease of cross section with increasing alpha energy is 
consistent with the general trend of increasing Z with increasing alpha 
energy and qualitatively consistent with the trend predicted by 
multi-nucleon transfer models. In this case the production of lower 
energy activities, while having contributions of feeding from higher Z, 
will tend to be dominated by direct production.

\begin{figure}
\epsfig{file=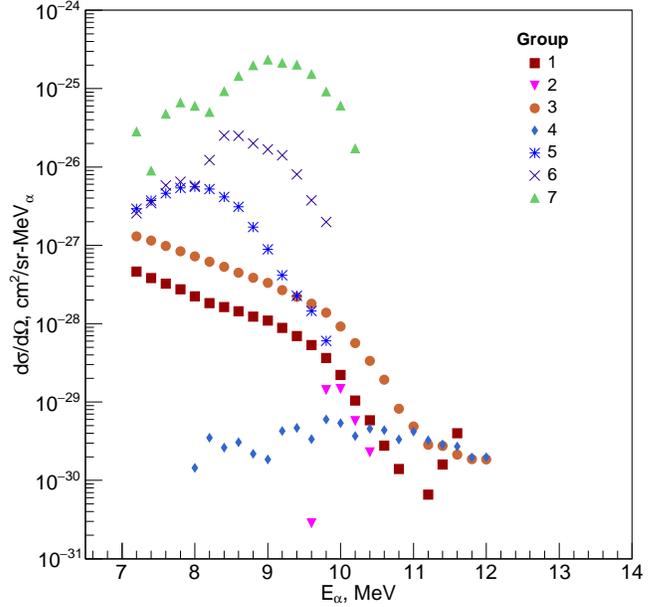,width=8.5cm,angle=0}
 \caption{Thick target differential cross setions, cm$^2$/sr-MeV in angular
range 31$^o$ - 50$^o$, see text.}
\label{fig12}
\end{figure}

The differential cross sections seen in figure~\ref{fig11} depend upon alpha 
energy, half-life and detection angle.  The mixture of activities in a given 
alpha energy range also can depend on pulsing protocol.  As the bulk of the 
data appear in the ring 2 portion of the active catcher we have chosen that 
ring, which spans an angular range 31$^o$ -50$^o$, for comparison of the 
differential cross sections for different half-life ranges.  As previously
 noted in figure~\ref{fig6},  7 different bands of sampled half-lives are 
observed.  We identified these, from top to bottom, as bands 1 through 7. 
In figure~\ref{fig12} we present the measured thick target differential 
cross sections for ring 2 for each of these bands. 

\section{Comparison to Previous Results}
In the late 70’s several groups employed similar multi-nucleon 
transfer reactions at energies ranging from the Coulomb barrier to 
8.5 MeV/nucleon to search for new elements from super heavy 
elements~\cite{hildenbrand77,gaggeler80,jungclas78,freisleben79,reidel79,kratzxx,schadelxx,trautmann78}.  These included both in-beam detection and 
radiochemical studies seeking evidence of new  spontaneously fissioning 
or alpha emitting nuclei. Both thin target and thick target irradiations 
were carried out. In all cases no new elements were observed and half-life 
dependent upper limits to heavy element production cross sections were 
reported.

The present data for thick target cross sections indicate cross sections 
which are somewhat in excess of those limits. It is natural, therefore, 
to ask why this is the case. For the previous radiochemical  and gas jet
 experiments thick targets were employed.  Time delays inherent in the 
radiochemical and jet techniques might account for some of the present 
differences.  Reference~\cite{schadelxx} also reports results of a rotating 
wheel collection experiment, but only to search for  spontaneous fission 
activities. We speculate that implantation depths of the products may have 
had some effect on the results reported.  

The previous experiment which may be most directly compared with ours is 
the in-beam experiment of references~\cite{hildenbrand77,freisleben79}. 
One significant difference is that their experiment employed a thin target 
so that a very small range of reaction energy at 7.42 MeV/u  was explored. 
In contrast our experiment explores the range from 7.5 MeV/u down to 
$\sim 6$ MeV/u.  Inspection of the alpha energy spectra in 
reference~\cite{freisleben79} reveals low level high energy signals which 
could be candidates for heavier element decay but were discounted because 
the microsecond time resolutions in the experiments did not allow sufficient 
discrimination against pile-up events. The alpha spectrum presented in 
reference~\cite{jungclas78} also shows some potentially interesting alpha 
particles below 11.6 MeV.  For energy above that the observed signals from 
two experiments for a total beam time of 5.5 hours indicate pile-up 
contributions similar to those invoked in reference~\cite{freisleben79}.  
In the present experiment modern flash ADCs were operated in a mode which 
allowed $\sim 16$ ns time resolution, greatly reducing pile up possibilities. 
In addition, the recording of the individual detector signal traces allowed 
inspection of individual detector signals.  Our analysis was restricted to 
events with flash ADC signals separated by 40 to 100 ns. 

\section{Summary and Conclusions}
The present experimental results for a survey of the production of 
$\alpha$-particle decaying heavy nuclei in reactions of 
7.5-6.1 MeV/nucleon $^{238}$U + $^{232}$Th indicate the observation of a 
number of previously unreported alpha emitters with energies reaching as 
high as 12 MeV.  Comparisons of the energies and half-lives of these alpha 
emitters with known and predicted half-lives suggest that new activities 
with Z as high as 116, and perhaps higher, are being produced in these 
reactions.  First cross section estimates imply that the cross sections 
are significantly higher than estimated by many models employing 
statistical decay calulations.  This may reflect a confluence of several 
factors, i.e. shell effects leading to higher barriers and lower 
excitation energies of the relevant primary nuclei, the importance of 
microscopic fission dynamics and  beta decay feeding by neighboring nuclei. 
It is our hope that the present data provide an incentive and a basic road 
map for further work in this direction. This could include more narrowly 
focused experiments with such an active catcher array and/or with 
appropriately designed spectrometers~\cite{barbui06,schadel07}.  We 
believe that a much improved active catcher array with higher granularity, 
better energy resolution and linear energy response is realizable using 
single crystal diamond detectors and faster electronics.  Such a detector 
would allow the establishment of parent daughter relationships and searches
for even smaller production rates.

\section{Acknowledgements}
During the course of this research our colleague Z. Sosin died. He was 
a major contributor to the project and is greatly missed. We thank 
A. Staszczak, A. Afanasiev, V. Karpov. K. Zhao, S. Giuliani and 
V. Zagrebaev (deceased), all of whom provided us with detailed 
tables of their theoretical results.  We also would like to acknowledge 
very useful conversations and exchanges with S. Umar, H. Freiseleben. 
We thank the operations staff of the TAMU Cyclotron Institute for all 
of their efforts in support of this work.  This work was supported by the 
United States Department of Energy under Grant \# DE-FG03- 93ER40773 and by 
The Robert A. Welch Foundation under Grant \# A0330 as well as by the 
National Science Center in Poland, contract no. UMO-2012/04/A/ST2/00082.

\end{document}